# Wave-like Tunneling of Phonons Dominates Glass-like Thermal Transport in Quasi-1D Copper Halide CsCu$_2$I$_3$

Jiongzhi Zheng[1], Changpeng Lin[2,3], Chongjia Lin[4], Baoling Huang[4,5,6,*], Ruiqiang Guo[7,†], Geoffroy Hautier[1,‡]

[1]Thayer School of Engineering, Dartmouth College, Hanover, New Hampshire, 03755, USA
[2]Theory and Simulation of Materials (THEOS), École Polytechnique Fédérale de Lausanne, CH-1015 Lausanne, Switzerland
[3]National Centre for Computational Design and Discovery of Novel Materials (MARVEL), École Polytechnique Fédérale de Lausanne, CH-1015 Lausanne, Switzerland
[4]Department of Mechanical and Aerospace Engineering, The Hong Kong University of Science and Technology, Clear Water Bay, Kowloon, Hong Kong
[5]HKUST Foshan Research Institute for Smart Manufacturing, Hong Kong University of Science and Technology, Clear Water Bay, Kowloon, Hong Kong, China
[6]HKUST Shenzhen-Hong Kong Collaborative Innovation Research Institute, Futian, Shenzhen 518055, China
[7]Thermal Science Research Center, Shandong Institute of Advanced Technology, Jinan, Shandong Province, 250103, China

## Abstract

Fundamental understanding of thermal transport in compounds with ultra-low thermal conductivity remains challenging, primarily due to the limitations of conventional lattice dynamics and heat transport models. In this study, we investigate the thermal transport in quasi-one-dimensional (1D) copper halide CsCu$_2$I$_3$ by employing a combination of first principles-based self-consistent phonon calculations and a dual-channel thermal transport model. Our results show that the 0-K unstable soft modes, primarily dominated by Cs and I atoms in CsCu$_2$I$_3$, can be anharmonically stabilized at ~ 75 K. Furthermore, we predict an ultra-low thermal conductivity of 0.362 Wm$^{-1}$K$^{-1}$ along the chain axis and 0.201 Wm$^{-1}$K$^{-1}$ along cross chain direction in CsCu$_2$I$_3$ at 300 K. Importantly, we find that an unexpected anomalous trend of increasing cross-chain thermal conductivity with increasing temperature for CsCu$_2$I$_3$, following a temperature dependence of ~$T^{0.106}$, which is atypical for a single crystal and classified as an abnormal glass-like behavior. The peculiar temperature-dependent behavior of thermal conductivity is elucidated by the dominant wave-like tunnelling of phonons in thermal transport of CsCu$_2$I$_3$ along cross-chain direction. In contrast, particle-like phonon propagation primarily contributes to the chain-axis thermal conductivity across the entire temperature range of 300-700 K. The sharp difference in the dominant thermal transport channels between the two crystallographic directions can be attributed to the unique chain-like quasi-1D structure of CsCu$_2$I$_3$. Our study not only illustrates the microscopic mechanisms of thermal transport in CsCu$_2$I$_3$ but also paves the way for searching for and designing materials with ultra-low thermal conductivity.

---

[*] mebhuang@ust.hk
[†] ruiqiang.guo@iat.cn
[‡] Geoffroy.t.f.hautier@dartmouth.edu

# I. INTRODUCTION

The effective management and utilization of thermal energy are vital for enhancing the operational efficiency and prolonging the lifetime of thermal data storage devices [1], photovoltaic [2-5] and optoelectronic devices [6], along with thermoelectric systems [7-9]. Halide perovskites have shown attractive prospects for photovoltaics and optoelectronics due to their outstanding properties, including a tunable band gap, high absorption coefficients, large carrier mobilities, long diffusion lengths, and exceptional defect tolerance [10,11]. Heat dissipation via phonon thermal transport and the lattice dynamics have a significant impact on the thermal stability, carrier lifetime, and optoelectronic performance of crystalline halide perovskites [12-14]. Furthermore, halide perovskites have recently gained interest in thermoelectrics for thermal-to-electric energy conversion, thanks to their advantageous electrical properties [15-17] and minimal thermal conductivity [4,5,18,19]. To enhance the efficiency of thermoelectric conversion, a central focus in thermoelectric materials research has been to reduce irreducible thermal transport while simultaneously maintaining exceptional electrical transport properties [8,20]. Many halide perovskites have been observed to exhibit remarkably low thermal conductivities in experimental measurements [2,4,5,21,22], owing to their soft lattice structures and strong anharmonicity. For instance, Pisoni *et al.* [21] first reported an ultra-low room-temperature thermal conductivity of 0.5 $Wm^{-1}K^{-1}$ in hybrid inorganic-organic halide perovskite $CH_3NH_3PbI_3$, which was attributed to the dynamic cation disorder and an expected soft lattice. Lee *et al.* [4] experimentally reported an ultra-low thermal conductivity of 0.45 and 0.42 $Wm^{-1}K^{-1}$ for all-inorganic halide perovskite $CsPbI_3$ and $CsPbBr_3$, respectively, which can be traced back to the highly anharmonic cluster rattling. Acharyya *et al.* [5] first experimentally observed an ultra-low thermal conductivity of ~0.37~0.28 $Wm^{-1}K^{-1}$ in the temperature range of 295~523 K, which is originated from the anharmonic soft vibrations of sublattice. Although the significant research progress of physical properties in halide perovskites, the environmental toxicity of the Pb ion and poor thermal and chemical stabilities remain a limitation for large-scale commercialization for all-inorganic lead halide perovskites [23] and organic-inorganic halide perovskites [24], respectively.

Recently, nontoxic and highly stable low-dimensional Cu halides have been attracting growing interest due to their distinctive physical properties such as high photoluminescence quantum efficiencies and widely adjustable emission wavelengths [25,26]. Prominent instances encompass $Cs_3Cu_2X_5$ (quasi-zero dimensional (0D)) [27,28], $Rb_2CuX_3$ (quasi-1D) [29], and $CsCu_2X_3$ [26,30,31] (quasi-1D), with X representing Cl, Br, and I. Extensive research has been conducted on the favorable optical characteristics of these metal halides [26,30-33], which are primarily attributed to the presence of robust self-trapped excitons resulting from their reduced dimensionality [34]. Despite considerable efforts on opto-electronic properties of low-dimensional metal halides, the lattice dynamics and thermal transport impacting their optoelectronic properties are yet to be further explored. Theoretical investigations into the thermal conductivity of $Cs_3Cu_2X_5$ (quasi-0D) [35] and $CsCu_2X_3$ [36] have been carried out, revealing remarkably low thermal conductivity values of 0.018 and 0.05 $Wm^{-1}K^{-1}$ at room temperature, respectively. However, these ultralow thermal conductivities were calculated using the lowest order of perturbation theory, considering only three-phonon scattering processes, a single particle-like phonon propagation channel, and 0-K harmonic phonons. Lattice vibrations in functional materials often display strong anharmonicity [37]. Consequently, the quasi-harmonic phonon theory, which doesn't account for temperature-dependent anharmonic interactions, fails to capture their lattice dynamics, particularly for opto-electronic and photovoltaic materials [19,36-38]. In these scenarios, it is essential to go beyond three-phonon scattering processes, as higher-order phonon scattering can play a significant role in determining the lattice thermal conductivity in highly anharmonic compounds. This phenomenon is evident in materials such as AgCrSe$_2$ [39], Cu$_{12}$Sb$_4$S$_{13}$ [40], and Tl$_3$VSe$_4$ [41]. Furthermore, materials exhibiting ultra-low thermal conductivity approaching the glass limit may challenge the effectiveness of the conventional Peierls-Boltzmann transport equation (PBTE) method [42] for explaining thermal transport [4,43-45]. In such cases, both the wave-like and particle-like characteristics of phonons may play pivotal roles in the thermal transport process [40,43-47]. Therefore, a thorough comprehension of the lattice dynamics and the microscopic mechanisms governing thermal transport in low-dimensional metal halides remains largely uncharted territory, presenting an urgent need for potential thermoelectric applications.

In this work, we thoroughly investigate the anharmonic lattice dynamics and the microscopic mechanisms of thermal transport in the quasi-1D copper halide CsCu$_2$I$_3$ by using a state-of-the-art first-principles scheme. Within this first-principles-based framework, we combine the anharmonic phonon renormalization technique and a unified theory of thermal transport to calculate the lattice thermal conductivity of CsCu$_2$I$_3$. To accurately capture the lattice dynamics and thermal transport characteristics, both the 3ph and 4ph scatterings are considered in self-consistent phonon calculations and dual-channel thermal transport mode, i.e., particle-like phonon propagation and wave-like tunnelling of phonons. At zero-K phonon calculations, we observe that the CsCu$_2$I$_3$ structure with a *cmcm* space group exhibits dynamical instability. However, it can be anharmonically stabilized at ~75 K, suggesting that it represents a low-temperature average structure. Furthermore, we demonstrate that the unique chain-like 1D structure leads to low-lying flat modes in mid-frequency region, consequently resulting in dramatically strong 4ph scattering rates in crystalline CsCu$_2$I$_3$. Using the dual-channel transport model, we show the effect of 4ph scatterings on particle-like phonon propagation and wave-like tunnelling of phonons and predicted an ultra-low thermal conductivity for CsCu$_2$I$_3$. We find that the conventional phonon-gas model fails to explain the thermal transport in cooper halide CsCu$_2$I$_3$. Particularly, the cross-chain thermal conductivity exhibits an abnormal upward trend with increasing temperature and this behavior is primarily attributed to the dominance of wave-like tunnelling of phonons in cross-chain thermal transport. Our work reveals the anomalous heat conduction physics in crystalline CsCu$_2$I$_3$, which also helps to understand thermal transport of other quasi-1D materials with ultra-low thermal conductivity.

## II. COMPUTATIONAL DETAILS AND METHEDOLOGY

**First-principles calculations and compressive Sensing technique**

Density functional theory (DFT) calculations for the CsCu$_2$I$_3$ crystal were carried out using projector-augmented wave (PAW) [48] pseudopotentials, which were implemented in the Vienna Ab initio Simulation Package (VASP) [49]. PAW pseudopotentials were employed to treat the Cs($5s^2 5p^6 6s^1$), Cu($3d^{10} 4s^1$) and I($4d^{10} 5s^2 5p^5$) shells as valence electrons and the PBEsol [50] of the generalized gradient approximation (GGA) [51] was ultilized to the exchange-correlation functional in all calcultions. A kinetic energy cut-off value of 520 eV and a 10×10×8 Monkhorst-Pack electronic *k*-point grid were used for structural optimization of primitive cell. Tight force and energy convergence criterions were set to $10^{-5}$ eV·Å$^{-1}$ and $10^{-8}$ eV, respectively, for both structural relaxation and self-consistent DFT calculations. The fully optimized lattice consants (*a* = 10.063, b =13.082 Å, c = 6.104 Å) agree well with the experimentally reported values (*a* = 10.548 Å, b =13.173 Å, c = 6.097 Å) for crystal with a ***cmcm*** space group [26,52].

The 0-K harmonic (2$^{nd}$-order) interatomic force constants (IFCs) were extracted using the finite-displacement method [53], as implemented in Phonopy [54]. A 3×3×2 supercell containing 216 atoms was adopted to achive good convergence for phonon dispersions, wherein a 3×3×4

Monkhorst-Pack electronic *k*-point grid and a kinetic energy cutoff of 520 eV were used in static DFT calculations. To map out the potential energy surfaces (PESs) for specific phonon modes, the same supercell dimension and input setting of the static self-consistent DFT calcultion were employed.

To accurately and efficiently obtain the anharmonic interatomic force constants (IFCs), we opted for the Compressive Sensing Lattice Dynamics method (CSLD) [55] instead of the traditional finite-displacement approach [53]. In particular, we applied the compressive sensing technique [56] of CSLD to identify and collect the physically significant anharmonic IFCs from the extensive set of irreducible anharmonic IFCs, utilizing a constrained displacement-force dataset [57]. Using the random number method, we generated a set of 100 atomic structures directly from an equilibrium $3\times3\times2$ supercell. These structures were created along random directions, with a uniform displacement of 0.15 Å applied to all atoms. Subsequently, we conducted a DFT calculation for the 100 atomic structures to acquire the atomic forces. In this calculation, we employed a $3\times3\times4$ Monkhorst-Pack electronic k-point grid and set the energy convergence criterion at $10^{-8}$ eV. In the final step, we utilized the displacement-force datasets and the 0-K harmonic interatomic force constants (IFCs) obtained to extract anharmonic IFCs up to the fourth order. This extraction was accomplished using the least absolute shrinkage and selection operator (LASSO) technique [58]. For the cubic and quartic IFCs extraction, we applied real-space cutoff radii of 7.41 Å and 5.30 Å, respectively. In this work, the IFCs estimation was performed by using the **ALAMODE** package [57,59] and our in-house code **Pheasy** [60].

**Anharmonic phonon energy renormalization technique**

We compute the anharmonically renormalized phonon energy at finite temperatures using the self-consistent phonon approximation (SCP) [57,61,62]. When considering only the first-order contribution to the phonon self-energy from quartic anharmonicity, the resulting diagonal SCP equation is expressed as follows:

$$\Omega_q^2 = \omega_q^2 + 2\Omega_q I_q, \tag{1}$$

where $\omega_q$ is the zero-K phonon frequency of phonon mode $q$ obtained from harmonic approxiamtion, and $\Omega_q$ is the renormalized phonon frequency including temperature effects. The quantity $I_q$ is defined as

$$I_q = \frac{1}{8N}\sum_{q'}\frac{\hbar V^{(4)}(q;-q;q';-q')}{4\Omega_q\Omega_{q'}}[1+2n(\Omega_{q'})], \tag{2}$$

where $N$ is the total number of sampled wave vectors, $\hbar$ is the reduced Planck constant, $n$ is the phonon popalation from the Bose-Einstein distribution and $V^{(4)}(q;-q;q';-q')$ is the reciprocal representation of fourth-order IFCs. Additionally, to accurately calculate the anharmonically renormalized phonon energy, we also include the off-diagonal terms of the phonon self-energy, i.e., polarization mixing (PM) [57], into the phonon renormalization calculation. In this study, we employed a $q$-mesh of 3×3×2 for the SCP calculations, which aligns with the dimensions of a 3×3×2 supercell utilized in the real-space-based phonon energy renormalization technique. Additionally, the SCP $q$-mesh of 3×3×2 was chosen to match the supercell dimensions relevant to the extraction of second-order IFCs in real space.

The anharmonically renormalized phonon energies, resulting exclusively from quartic anharmonicity, may be overestimated for highly anharmonic materials such as BaZrO$_3$ [63], Cu$_{12}$Sb$_4$S$_{13}$ [40] and Cs$_2$AgBiBr$_6$ [38]. As a result, we also take into account the negative energy shifts from bubble diagram due to cubic anharmonicity in addition to the above renormalized phonon energies. The resulting SCP calculation, enhanced by the inclusion of the bubble diagram correction and referred to as SCPB, is defined as [37]

$$\left(\Omega_q^B\right)^2 = \Omega_q^2 - 2\Omega_q \operatorname{Re} \Sigma_q^B [G, \Phi_3](\Omega = \Omega_q^B), \tag{3}$$

where $B$ stands for the bubble diagram, $\Phi_3$ is the third-order IFCs explicitly included in the anharmonic self-energy calculation, and $\Sigma_q^B[G, \Phi_3](\Omega_q)$ is the frequency-dependent phonon bubble self-energy. In this study, the anharmonic phonon renormalization was performed using the **ALAMODE** package [37,57]. Within the framework of the quasiparticle (QP) approximation, several treatment options, including QP[0], QP[S], and QP-NL, can be selected to solve Eq. (3) and obtain the fully renormalized phonon energies accounting for cubic and quartic anharmonicities [37]. Following the recommendation in the literature [37], we opted for the QP-NL treatment to evaluate the negative phonon energy shifts resulting from the bubble diagram. Furthermore, we ignore the minor phonon energy shifts caused by the tadpole diagram from cubic anharmonicity, a phenomenon observed in many crystals [40,63,64]. It's worth noting that in this study, we utilized our in-house code for the transformation of force constants between the **ALAMODE** [57,59] and **ShengBTE** [65] packages.

**Intrinsic and Extrinsic Phonon Scattering Rates**

The intrinsic multi-phonon scattering rates, including both three-phonon (3ph) $\Gamma_q^{3ph}$ and four-phonon (4ph) $\Gamma_q^{4ph}$ scattering rates, can be derived from Fermi's golden rule from perturbation theory [66]. Under the single-mode relaxation time approximation (SMRTA) treatment, the $\Gamma_q^{3ph}$ and $\Gamma_q^{4ph}$ can be expressed as [40,41,66]

$$\Gamma_q^{3\text{ph}} = \sum_{q'q''} \left\{ \frac{1}{2}\left(1 + n_{q'}^0 + n_{q''}^0\right)\zeta_- + \left(n_{q'}^0 - n_{q''}^0\right)\zeta_+ \right\}, \tag{4}$$

$$\Gamma_q^{4\text{ph}} = \sum_{q'q''q'''} \left\{ \frac{1}{6}\frac{n_{q'}^0 n_{q''}^0 n_{q'''}^0}{n_q^0}\zeta_{--} + \frac{1}{2}\frac{\left(1 + n_{q'}^0\right)n_{q''}^0 n_{q'''}^0}{n_q^0}\zeta_{+-} + \frac{1}{2}\frac{\left(1 + n_{q'}^0\right)\left(1 + n_{q''}^0\right)n_{q'''}^0}{n_q^0}\zeta_{++} \right\}, \tag{5}$$

where the phonon mode $q$ is a composite index of the wavevector $q$ and phonon branch $j$, and $n$ is the phonon popalation from the Bose-Einstein distribution, with

$$\zeta_{\pm} = \frac{\pi\hbar}{4N}\left|V^{(3)}(q,\pm q',-q'')\right|^2 \Delta_{\pm} \frac{\delta(\Omega_q \pm \Omega_{q'} - \Omega_{q''})}{\Omega_q \Omega_{q'} \Omega_{q''}}, \qquad (6)$$

and

$$\zeta_{\pm\pm} = \frac{\pi\hbar^2}{8N^2}\left|V^{(4)}(q,\pm q',\pm q'',-q''')\right|^2 \Delta_{\pm\pm} \frac{\delta(\Omega_q \pm \Omega_{q'} \pm \Omega_{q''} - \Omega_{q'''})}{\Omega_q \Omega_{q'} \Omega_{q''} \Omega_{q'''}}, \qquad (7)$$

Where $V^{(3)}(q,\pm q',-q'')$ and $V^{(4)}(q,\pm q',\pm q'',-q''')$ are the reciprocal representation of 3$^{\text{rd}}$- and 4$^{\text{th}}$-order IFCs, respectively [65,67], the delta function $\delta(\Omega)$ is enforced to ensure energy conservation in both 3ph and 4ph scatterings, and the Kronecker deltas $\Delta_{\pm}$ and $\Delta_{\pm\pm}$ are enforced to ensure momentum conservation for 3ph and 4ph scattering processes, respectively.

The extrinsic phonon scatterings from isotopes, denoted as $\Gamma_q^{isotope}$, can be expressed as [68]

$$\Gamma_q^{\text{isotope}} = \frac{\pi\Omega_q^2}{2N}\sum_{i\in u.c.} g(i)\left|e_q^*(i)\cdot e_{q'}(i)\right|^2 \delta(\Omega-\Omega'), \qquad (8)$$

where $e_q(i)$ is the eigenfunction of phonon mode $q$ at the atom $i$, and $g(i)$ is the Pearson deviation coefficient and its detailed expression can be referred to literature [68].

By combining the intrinsic and extrinsic phonon scattering rates and applying Matthiessen's rule, the total phonon scattering rate $\Gamma_q$ can be expressed as

$$\Gamma_q = \Gamma_q^{3\text{ph}} + \Gamma_q^{4\text{ph}} + \Gamma_q^{\text{isotope}}, \qquad (9)$$

**Unified Theory of Thermal Transport**

We use a unified theory of thermal transport [46,47] to compute the total lattice thermal conductivity $\kappa_{\text{L}}$. The total $\kappa_{\text{L}}$ can be obtained by summing the populations' contribution $\kappa_{\text{L}}^{\text{P}}$ and coherences' contribution $\kappa_L^c$, namely $\kappa_L = \kappa_L^P + \kappa_L^C$, where $\kappa_L^P$ and $\kappa_L^C$ account for the diagonal and off-diagonal terms of heat flux operators, respevtively. Specifically, $\kappa_L^P$ and $\kappa_L^C$ arise from the particle-like phonon propagation and wave-like tunnelling of phonons, respectively. The detailed formula for $\kappa_L^{P/C}$ under the SMRTA can be expressed as [46,47]

$$\kappa_L^{P/C} = \frac{\hbar^2}{k_B T^2 VN} \sum_q \sum_{j,j'} \frac{\Omega_{qj} + \Omega_{qj'}}{2} \upsilon_{qjj'} \otimes \upsilon_{qj'j} \cdot \frac{\Omega_{qj} n_{qj}(n_{qj}+1) + \Omega_{qj'} n_{qj'}(n_{qj'}+1)}{4(\Omega_{qj} - \Omega_{qj'})^2 + (\Gamma_{qj} + \Gamma_{qj'})^2} (\Gamma_{qj} + \Gamma_{qj'}), \quad (10)$$

where $k_B$, $T$, $V$ and $\upsilon$ are, respectively, the Boltzmann constant, the temperature in Kelvin, the primitive-cell volume and the group velocity matix, including both the diagonal and off-diagonal terms [69]. In Eq.(10), when $j = j'$, it results in the populations' contribution (PBTE result) $(\kappa_L^P)$, otherwise, it gives the coherences' contribution ($\kappa_L^c$). In solving Eq. (10), we used a $q$ mesh of 9×9×13 for the 3ph and 4ph scattering processes, with scalebroad parameters set at 0.5 and 0.02, respectively. These parameter choices resulted in well-converged results for the CsCu$_3$I$_2$ crystal. Additionally, the iterative scheme for the PBTE is exclusively applied to handle 3ph scattering processes. In contrast, the treatment of 4ph scattering processes is conducted at the SMRTA level due to the exceptionally high memory demands [40,67]. In this work, thermal conductivity calculations, including populations' and coherences' contributions, were carried out using the **ShengBTE** [65] and **FourPhonon** [67] packages, and our in-house code [38,63].

## III.   RESULTS AND DISCUSSIONS

### a) Crystal Structure, Phonon Dispersions and Lattice Anharmonicity

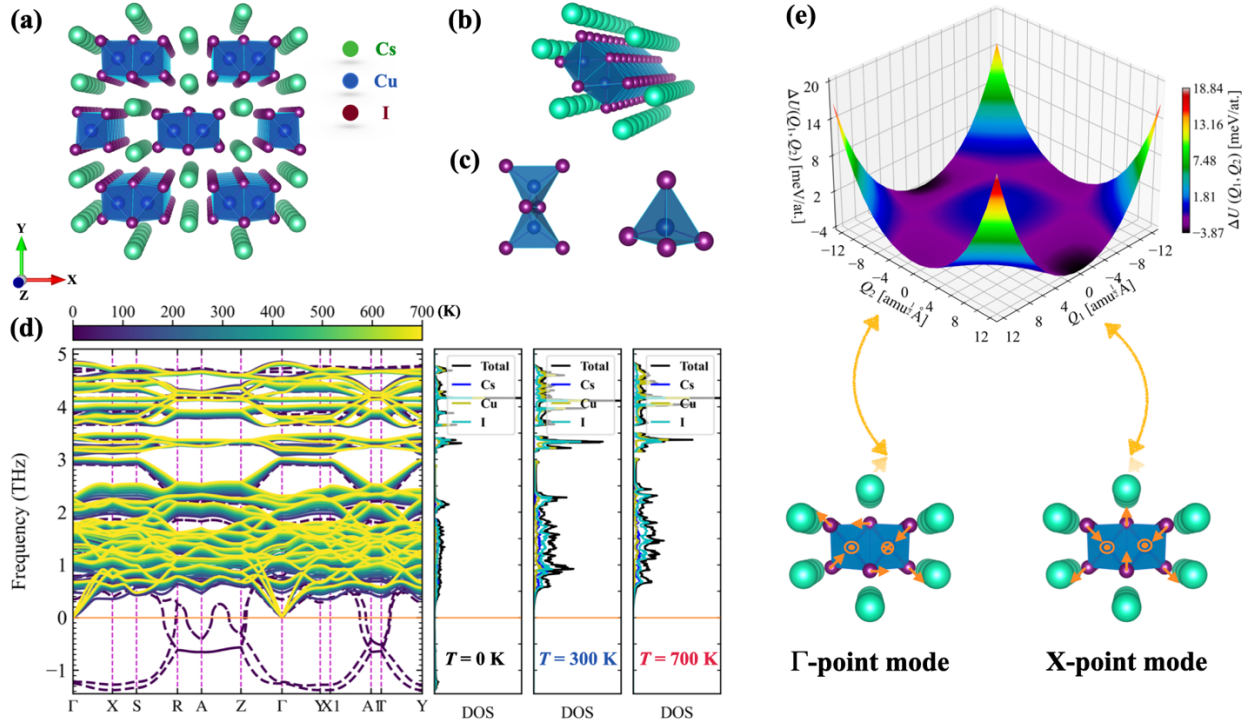

FIG. 1. (a) Crystal structure of cubic cooper halide $CsCu_2I_3$, which features one-dimensional (1D) Cu-I chains surrounded by Cs atoms. (b) A perspective view of the 1D polyanionic $[Cu_2X_3]^-$ chain with loosely bounded Cs atoms, wherein the polyanionic $[Cu_2X_3]^-$ units are formed by edge-shared connection of $Cu_2X_3$ tetrahedra. (c) Polyanionic $[Cu_2X_3]^-$ polyhedron and $Cu_2X_3$ tetrahedra, respectively. (d) Calculated an-harmonically renormalized phonon dispersions at finite temperatures (from 100 to 700 K) compared with phonon dispersions computed by harmonic approximation treatment at 0 K. The calculated atom-decomposed partial and total phonon densities of states (DOS) at 0, 300 and 700 K are shown on the right panel, respectively. (e) Calculated two-dimensional (2D) potential energy surface of $CsCu_2I_3$ as a function of normal mode coordinates $Q_1$ and $Q_2$. Here the soft modes at the Γ and X points are associated with the vibrations of Cu and I atoms and these vibrational animations are denoted by arrows and in/out circles.

We start by analyzing the anharmonic lattice dynamics of the 1D $CsCu_2I_3$ crystal structure at finite temperatures, as illustrated in Fig. 1. In Figures 1(a-c), $Cu^+$ ions occupy the interstitial sites in the tetrahedrally coordinated sublattice formed by $I^-$ anions, resulting in the formation of $[CuI_4]^{3-}$ units. The paired $[CuI_4]^{3-}$ tetrahedral strings, nestled within rhombic columnar cages formed by Cs+ cations, stack infinitely via edge sharing, giving rise to the $[Cu_2I_3]^-$ anionic chains of $CsCu_2I_3$, as shown in Figs. 1(b-c) [26,32]. Furthermore, previous study had revealed that the orbitals of Cu and I atoms predominantly influence the valence and conduction band minimum, indicating that Cs+ cations serve as a quasi-isolating shell for the 1D $[Cu_2I_3]^-$ chains [32].

Next, we calculate the vibrational properties of the 1D cooper halide $CsCu_2I_3$ using the harmonic approximation (HA) approach [42]. In Fig. 1(d), phonon softening is evident across the entire Brillouin zone, indicating the dynamical instability of $CsCu_2I_3$ at low temperatures, a result consistent with prior calculation [36]. The soft modes predominately arise from both Cu and I atoms, as evidenced by the atom-decomposed partial DOS in the right panel of Figure 1(d) and the projected atomic participation ratio (see Fig. S1(a-d) in the supplementary material (SM) [70]). To account for temperature effect on phonons in $CsCu_2I_3$, we consider both cubic and quartic anharmonicities in the phonon energy renormalization. Particularly, the cubic anharmonicity contributes to the negative phonon energy shifts, which is found to be crucial for predicting anharmonic phonon behavior in highly anharmonic compounds [37,40,71,72]. After anharmonic phonon renormalization, we observe a substantial stiffening of the soft modes with increasing temperature, as shown in Fig. 1(d). Furthermore, it's noteworthy that all anharmonically renormalized phonons exhibit stabilization above 75 K, as illustrated in Fig. S2 (a-c) of the SM [70]. This relatively low phase transition temperature provides an explanation for the experimentally observed $CsCu_2I_3$ structure, characterized by a *cmcm* space group at room temperature [26,32,52]. In addition to the Cu/I-dominated soft modes, the low-frequency optical modes (≤ 2.5 THz), mainly contributed by I atoms, also experience a progressive stiffening [see Figs. 1(d) and S1(a,c) in the SM [70]]. In the low-frequency regime, phonon modes primarily dominated by Cs atoms exhibit behavior similar to the rattler-like modes observed in caged compounds [61,73,74]. The rattling motion of these modes is demonstrated by the small atomic participation ratio and their temperature-dependent phonon stiffening is attributed to the large mean-square atomic displacements (MSD) of Cs atoms [see Figs. S3-4 in the SM [70]]. Conversely, the high-frequency phonon modes (> 2.5 THz) in $CsCu_2I_3$ are primarily dominated by the relatively light Cu atoms, as depicted in Figs. S1(a) and (d) in the SM [70]. These modes display a minor or negligible temperature-dependent stiffening, akin to the high-frequency phonon behavior observed in compounds like $BaZrO_3$ [63] and others [64].

To gain a more profound understanding of lattice anharmonicity and instability in the $CsCu_2I_3$ crystal, we construct a 2D potential energy surface (PES) [75] for the soft modes at the Γ and X points in Fig. 1(e). As expected, the minimum energy is situated beyond the zero-tilt amplitude ($Q_1 = Q_2 = 0$) for both soft modes at the Γ and X points, respectively, indicating strong lattice anharmonicity and dynamical instability [see Fig. 1(e)]. The soft modes at the Γ and X points are exclusively linked to the lattice vibrations of the 1D $[Cu_2I_3]^-$ chains, despite the fact that the Cs atoms exhibit the largest MSD [see Figs. 1(e) and S2 of the SM [70]]. In this context, the out-of-plane motions are attributed to I atoms, while the in-plane motions are assigned to Cu atoms, as illustrated in Fig. 1(e). Evidently, the source of lattice instability in $CsCu_2I_3$ differs from that in double perovskite [19,38], where the rotational sublattice causes the lattice instability. Although the minimum energies are located beyond the zero-tilt amplitude for both soft modes, it's important to highlight that these minima are relatively shallow, predicting approximately -2.57 and -3.91 meV/atom, respectively. This observation suggests that the high-temperature phase of $CsCu_2I_3$ crystal effectively represents an averaged structure at relatively low temperatures, analogous to compounds like $CsSnI_3$, $CsSnBr_3$ and $CsSnCl_3$ [76,77].

**b) Bond strength and phonon group velocity**

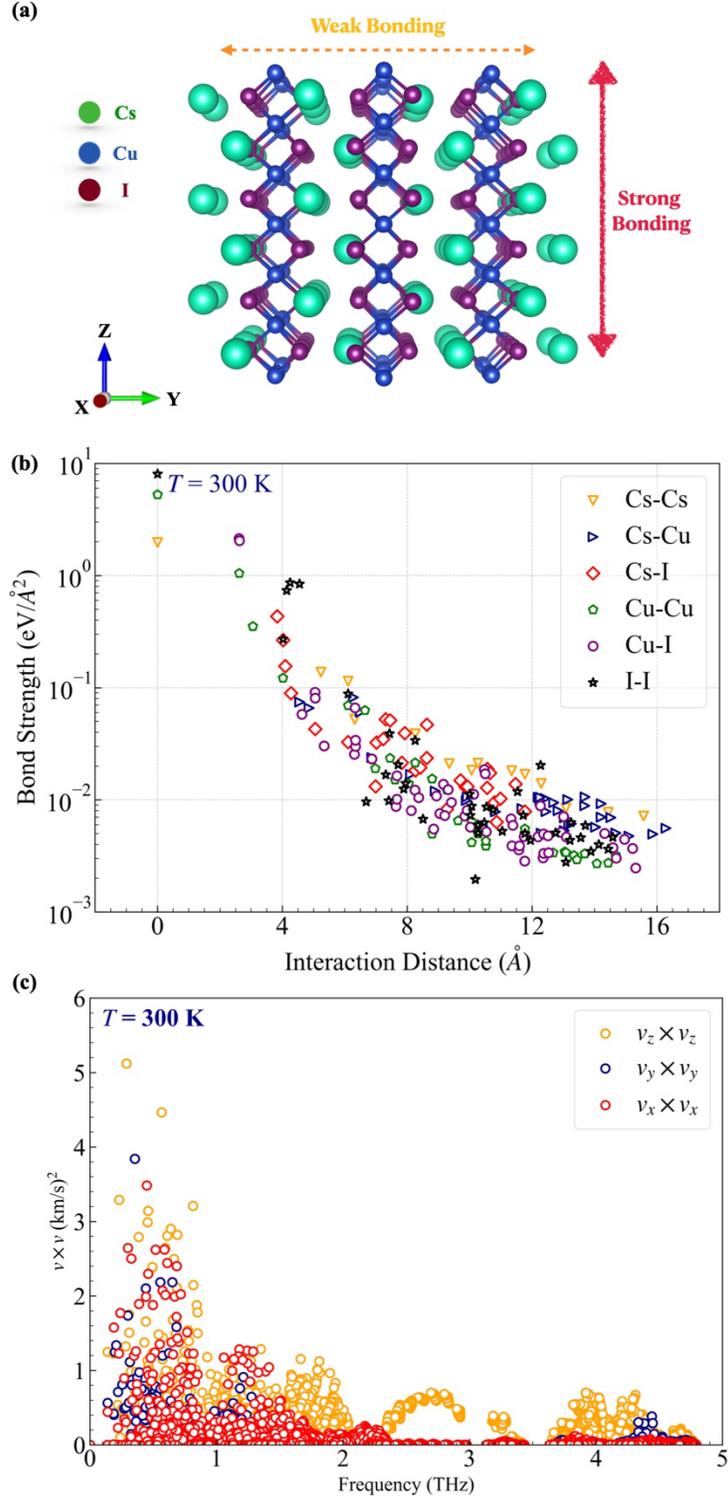

FIG. 2. (a) A perspective view of Cubic copper halide CsCu$_2$I$_3$ projected along the X axis. (b) Calculated bond strength as a function of interaction distance of atomic pairs at 300 K. (c) Calculated phonon group velocities along x, y and z axis, respectively, using the finite-temperature second-order interatomic force constants at 300 K.

Considering the crucial role of bond strength in characterizing thermal transport in compounds [78,79] we now turn our attention to the examination of bond strengths and group velocities in crystalline $CsCu_2I_3$. Given the unique chain-like structure of $CsCu_2I_3$, we expect strong bonding features along the chain axis, contrasting with weak bonds in the cross-chain directions, as illustrated in Fig. 2(a). From figure 2(b), it is evident that the in-chain bond strengths of nearest-neighbor Cu-I and Cu-Cu pairs exhibit highest values of ~2 and 1.05 eV/Å², respectively. In contrast, small values of ~0.4 and 0.08 eV/Å² are observed for the cross-chain bond strengths of nearest-neighbor Cs-I and Cs-Cu pairs, respectively. The weak interaction between Cs and I/Cu atoms is a key factor contributing to the rattling vibrations of Cs atoms in $CsCu_2I_3$, consequently leading to small group velocity and strong anharmonicity [5,61,73,80]. Furthermore, the significant differences between the in-chain and cross-chain bond strengths are expected to result in strong anisotropy in group velocity. Indeed, the group velocities along the chain axis (Z-axis) are consistently greater than those in the cross-chain directions (X/Y axes) [see Fig. 2(c)], thus giving rise to anisotropic thermal conductivity. Importantly, in comparison to the group velocities along the chain-axis (Z-axis), the cross-chain (X/Y-axis) group velocities exhibit rapid decay, particularly for phonons with frequencies exceeding 2 THz. We attribute this anomalous phenomenon to the 'quiescence' behavior of $Cs^+$ ion, as evidenced by the PDOS and atomic participation ratio [see Figs. 1(d) and S1(a-d) in the SM [70]]. More specifically, Cs atoms do not participate in the collective vibrations of atoms with frequencies exceeding 2 THz, which impedes the phonon propagation along cross-chain directions [71,81]. The exceptionally low cross-chain group velocities for high-frequency phonons (> 2 THz) lead to the loss of the phonon quasi-particle picture, thus enhancing the nature of wave-like tunnelling of phonons. Consequently, the distinct features between chain-axis and cross-chain group velocities may give rise to different mechanisms of thermal transport along those orientations.

## c) Phonon scattering rates, phase spaces and lifetimes

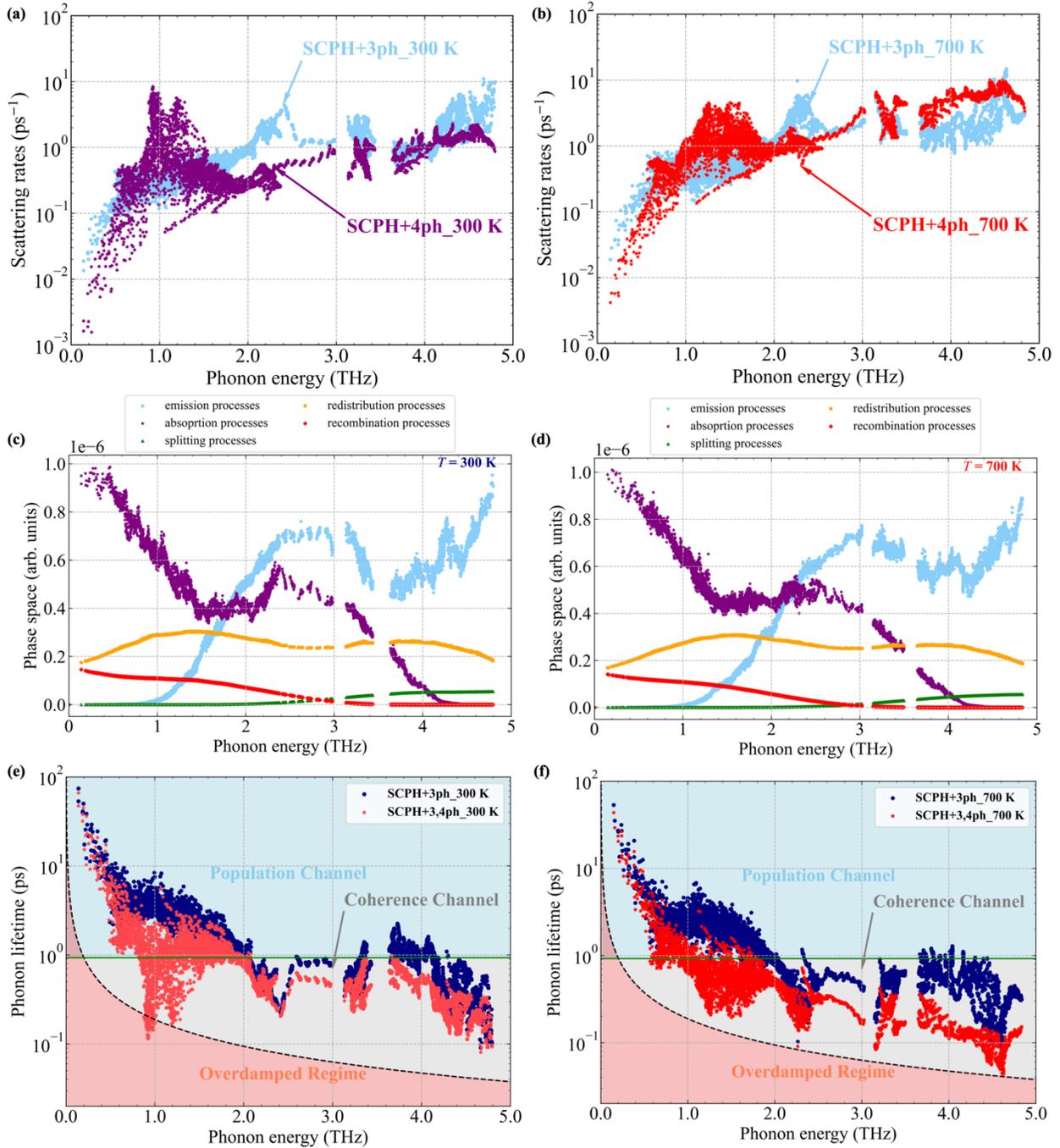

FIG. 3. (a) Calculated three- (3ph) and four-phonon (4ph) scattering rates at 300 K using anharmonically renormalized phonon frequencies and eigenvectors. (b) The same as (a), but at 700 K. (c) Calculated scattering phase spaces, including emission and absorption processes for three-phonon scatterings as well as splitting, redistribution and recombination processes for four-phonon scatterings at 300 K. (d) The same as (c), but at 700 K. (e) Calculated phonon lifetimes for three-phonon scattering processes, as well as for both the three- and four-phonon scattering processes at 300 K using anharmonically renormalized phonon frequencies and eigenvectors. The horizontal green solid line represents the Wigner limit in time [47]. Phonons with a lifetime above this limit primarily contribute to the populations' thermal conductivity $\kappa_L^P$, while those below mainly contribute to coherences' thermal conductivity $\kappa_L^C$. The dashed black curve depicts the Ioffe-Regel limit in time [82]. Phonons with a lifetime larger than the Ioffe-Regel limit are considered well-defined, whereas those below are recognized as overdamped. (f) The same as (e), but at 700 K.

Taking into account the strong anharmonicity in CsCu$_2$I$_3$ crystal, we explicitly incorporate both cubic and quartic anharmonicities into phonon scattering rates for the determination of thermal conductivity $\kappa_L$ [42,67]. In Figs. 3(a) and 3(b), we compute the 3-phonon (3ph) and 4-phonon (4ph) scattering rates by using anharmonically renormalized phonon modes alongside the 3$^{rd}$ and 4$^{th}$ force constants at temperatures of 300 and 700 K, respectively. We observe that at 300 K, $\Gamma_{4ph}$ surpasses $\Gamma_{3ph}$ for modes with frequencies in the 0.5 to 1.5 THz range, which shifts to 0.6–1.8 THz at 700 K [see Figs. 3(a-b)]. Obviously, the 4ph interaction processes not only play a crucial role in accurately predicting finite-temperature phonon energies but also determining thermal transport for CsCu$_2$I$_3$ crystal. Additionally, the $\Gamma_{4ph}$ within the mentioned frequency interval undergoes a sudden increase, while $\Gamma_{3ph}$ does not demonstrate a similar abrupt rise. Meanwhile, we observe a direct correlation between the peak of the 4ph scattering rates and the temperature-dependent shift in Cs-dominated phonons [see Figs. 1(d), 3(a-b) and S1(d)]. Hence, we attribute the abrupt rise in 4ph scattering rates to the presence of Cs-dominant rattling modes, i.e., low-lying flat modes. This phenomenon is not observed in frequency regions where Cs atoms are not involved [see Figs. 1(d), 3(a-b) and S1(a-d) in the SM [70]]. Similarly, the low-lying flat phonon modes consistently exhibit high phonon scattering rates in many crystals, including Cu$_{12}$Sb$_4$S$_{13}$ [40], Ba$_8$Ga$_{16}$Ge$_{30}$ [73], AgCrSe$_2$ [39] and YbFe$_4$Sb$_{12}$ [64,80]. The quasi-one-dimensional, chain-like structure of crystalline CsCu$_2$I$_3$ induces rattling vibrations of Cs atoms, consequently leading to the prevalence of dominant 4ph scattering rates. Interestingly, it's worth noting that the 3ph scattering phase spaces are generally larger than the 4ph phase spaces, implying that the significant 4ph scattering rates result from the extensive 4ph scattering matrices [see Figs. 3(c-d)]. This differs from observations in materials such as $Cu_{12}Sb_4S_{13}$ [40] and $Ba_8Ga_{16}Ge_{30}$ [73].

Recently, Simoncelli *et al.* [46,47] have introduced a novel parameter, i.e., the Wigner limit in time, for characterizing phonons in compounds with complex structures or/and strong anharmonicity. This parameter ais in categorizing phonons into distinct thermal transport regimes, where they are either dominated by particle-like phonon propagation or wave-like tunnelling of phonons channels. The Wigner limit in time is defined as $\tau = 3N_{at}/\omega_{max}$, where $N_{at}$ is the number of atoms in the primitive cell and $\omega_{max}$ is the maximum phonon frequency. In Figs. 3(e-f), when considering only 3ph scattering processes, it becomes apparent that most phonons with frequencies exceeding 2 THz are located below the Wigner limit in time. This observation underscores the importance of coherence channel (wave-like tunnelling of phonons) in thermal transport of crystalline $CsCu_2I_3$. Nonetheless, the lifetimes of phonons with frequencies below 2 THz, which serve as the principal heat carriers, still exceed the Wigner time limit. This suggests that perhaps the population channel, i.e., particle-like phonon propagation, continues to govern thermal transport. The inclusion of additional 4ph scattering processes leads to a significantly increased population of phonons with lifetimes lower than the Wigner limit in time [see Fig. 3(e-f)]. Notably, a substantial proportion of phonons with frequencies less than 2 THz dip below the Wigner time limit, some even approaching the overdamped region. This observation emphasizes the significance of the strong 4ph scatterings induced by the chain-like structure of $CsCu_2I_3$ in accurately unveiling the mechanisms of thermal transport. It also implies that the coherence channel probably holds a dominant role in thermal transport in crystalline $CsCu_2I_3$, as depicted in Figs. 3(e-f).

**d) Cumulative and Spectral thermal conductivity**

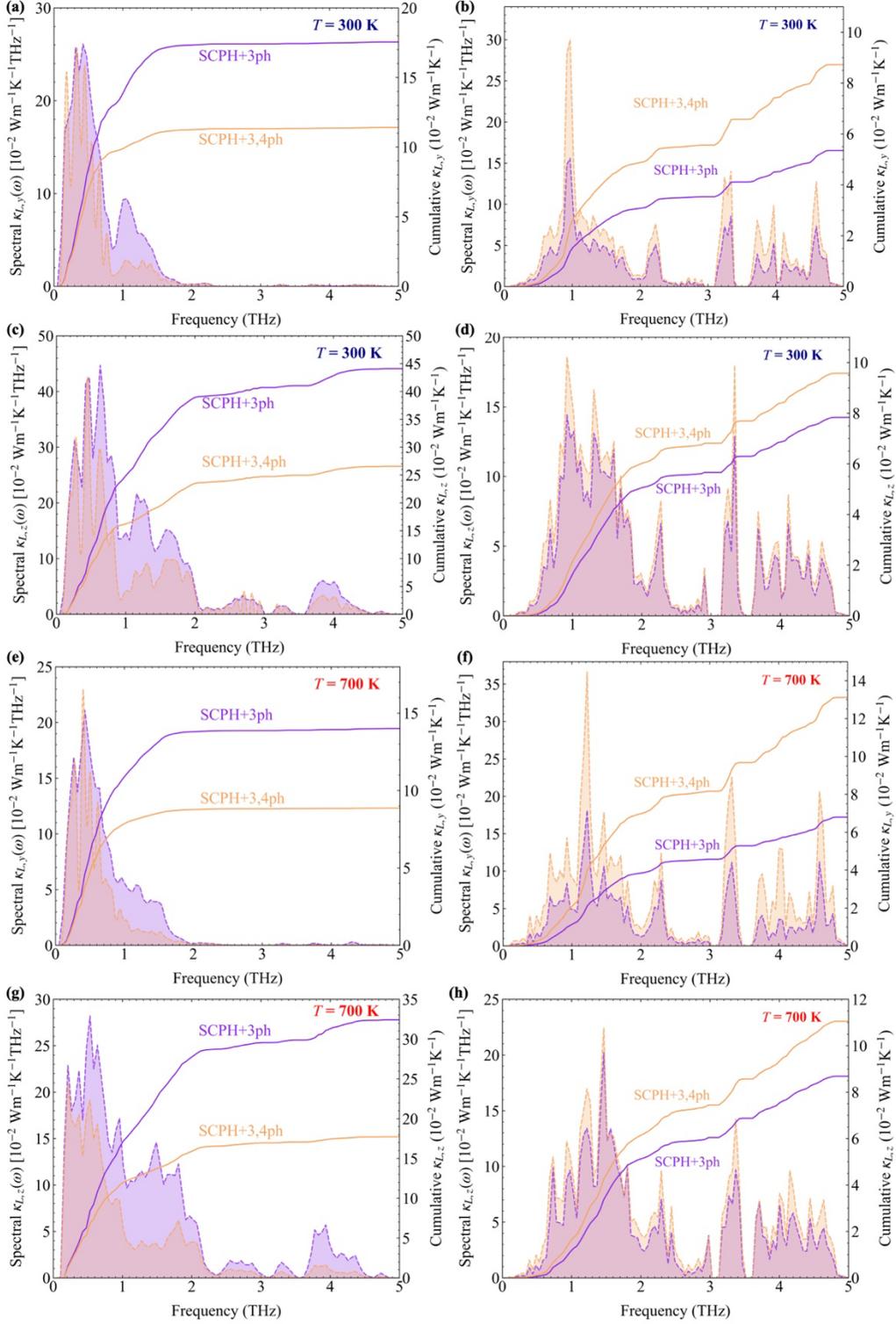

FIG. 4. Comparison of spectral/cumulative populations' and coherences' thermal conductivity obtained with (SCPH+3,4ph model) or without (SCPH+3ph model) 4ph scattering rates 300 and 700K, respectively. (a) Spectral and cumulative populations' thermal conductivity along the y axis obtained at 300 K. (b) Spectral and cumulative coherences' thermal conductivity along the y axis obtained at 300 K. (c) The same as (a), but along the z axis. (d) The same as (b), but along the z axis. (e) The same as (a), but at 700 K. (f) The same as (b), but at 700 K. (g) The same as (e), but along the z axis. (h) The same as (f), but along the z axis.

We move on to calculate the spectral and cumulative populations' thermal conductivity $\kappa_L^P$ and coherences' thermal conductivity $\kappa_L^C$ at 300 and 700 K, respectively, as illustrated in Figs. 4(a-h) and S5(a-b) in the SM [70]. Overall, on top of the thermal conductivities obtained considering only 3ph scattering processes, the inclusion of 4ph scattering processes leads to a decrease in $\kappa_L^P$ but an increase in $\kappa_L^C$ [40,63]. Furthermore, the chain-axis thermal conductivity $\kappa_{L,z}^P$ is approximately twice that of the cross-chain thermal conductivity $\kappa_{L,x}^P$, whether or not we consider 4ph scatterings, at both 300 K and 700 K [see Figs. 4(a, c, e, g)]. The remarkable anisotropy in particle-like phonon thermal transport observed in crystalline CsCu$_2$I$_3$ can be traced back to the unique chain-like 1D structure, which induces significant anisotropy in group velocity [see Fig. 2(c)]. On the contrary, a weak anisotropy is observed between the chain-axis $\kappa_{L,z}^C$ and cross-chain thermal conductivity $\kappa_{L,x}^C$ for crystalline CsCu$_2$I$_3$, as shown in Figs. 4(b, d, f, h).

We now delve deeper into a more comprehensive understanding of the spectral and cumulative contributions to thermal conductivity, namely $\kappa_L^P$ for populations and $\kappa_L^C$ for coherences. Figures 4(a, c, e, g) clearly illustrate that phonons with frequencies less than 2 THz are the primary heat carriers contributing to $\kappa_L^P$ in both the chain-axis and cross-chain directions. When considering only 3ph scattering processes, an ultra-low $\kappa_L^P$ of 0.175 (0.44) Wm$^{-1}$K$^{-1}$ and 0.14 (0.32) Wm$^{-1}$K$^{-1}$ is observed for the cross-chain (chain-axis) direction at 300 and 700 K, respectively. The ultra-low thermal conductivity in in CsCu$_2$I$_3$ is primarily attributed to its short phonon lifetimes induced by strong anharmonicity. This conclusion is drawn from a comparison of key factors, including group velocity and phonon lifetime [see Figs. 2(c) and 3(a-b)], between crystalline CsCu$_2$I$_3$ and Si [83], using the simple kinetic theory of thermal conductivity [79]. By further incorporating 4ph scattering rates, the corresponding predicted $\kappa_L^P$ decreases to 0.114 (0.266) Wm$^{-1}$K$^{-1}$ and 0.089 (0.177) Wm$^{-1}$K$^{-1}$ representing approximately a 35% (40%) and 36% (45%) reduction, respectively, as illustrated in Figs. 4(a, c, e, g). The remarkable reduction in $\kappa_L^P$ due to 4ph scatterings can be attributed to phonons falling within the frequency range of 0.5-2 THz, which aligns with the dominant region of 4ph scattering rates induced by chain-like structure of $CsCu_2I_3$ [see Figs. 3(a-b) and 4(a, c, e, g)].

For coherences' contributions to thermal conductivity $\kappa_L^C$, we predict a value of 0.053 (0.078) $Wm^{-1}K^{-1}$ and 0.068 (0.087) $Wm^{-1}K^{-1}$ along the cross-chain (chain-axis) direction at 300 and 700 K, respectively [see Figs. 4(b, d, f, h)]. These coherences' thermal conductivities account for 23.25% (15.06%) and 32.64% (21.17%) of the corresponding total thermal conductivities, namely, $\kappa_L = \kappa_L^C + \kappa_L^P$, respectively. Even though the $\kappa_L^C$ makes a significant contribution to the total $\kappa_L$, the $\kappa_L^P$ remains the dominant role in thermal transport in $CsCu_2I_3$ when considering only 3ph scattering processes. Incorporating the additional 4ph scattering effects, the $\kappa_L^C$ increases to 0.087 (0.096) $Wm^{-1}K^{-1}$ and 0.131 (0.111) $Wm^{-1}K^{-1}$ for the cross-chain (chain-axis) direction at 300 and 700 K, respectively. Consequently, the enhanced percentages are 64.15% (23%) and 92.65% (27.58%) at 300 and 700 K, respectively. In contrast to the cause of $\kappa_L^P$ reduction due to 4ph scatterings, the significant enhancement in $\kappa_L^C$ resulting from 4ph scattering is attributed not only to phonons within the frequency range of 0.5-2 THz but those in higher frequency ranges [see Figs. 4(b, d, f, h)]. Notably, the enhanced cross-chain $\kappa_{L,x}^C$ due to 4ph scatterings surpasses the corresponding reduced $\kappa_{L,x}^P$ due to 4ph scatterings at 700 K, as illustrated in Figs. 4(e) and (f). This observation indicates that the conventional Peierls-Boltzmann equation [42] is inadequate for accurately describing thermal transport in $CsCu_2I_3$, especially in the cross-chain direction. The dominance of $\kappa_{L,x}^C$ in the cross-chain thermal conductivity of $CsCu_2I_3$ can be traced back to its unique chain-like 1D structure. This unique structure, characterized by weak bonding, results in an exceedingly low cross-chain $\kappa_{L,x}^P$, exceptionally small cross-chain group velocities [see Fig. 2(c)], and strong 4ph scatterings [see Figs. 3(a-b)], all of which collectively facilitate the wave-like tunnelling of phonons. Figures 4(b, d, f, h) also reveal that 4ph scatterings have a more pronounced impact on cross-chain $\kappa_{L,x}^C$ when compared to chain-axis $\kappa_{L,z}^C$. This distinction can be attributed to the strong wave-like nature of phonons due to exceptionally low group velocities along the cross-chain direction. While $\kappa_{L,x}^C$ governs cross-chain thermal transport, $\kappa_{L,z}^P$ continues to dominate the total chain-axis thermal conductivity across the temperature range of 300-700 K. This observation suggests that particle-like phonon propagation governs chain-axis thermal transport, while wave-like tunnelling of phonons dominates cross-chain thermal transport in $CsCu_2I_3$.

# e) The Two-dimensional modal coherences' conductivity

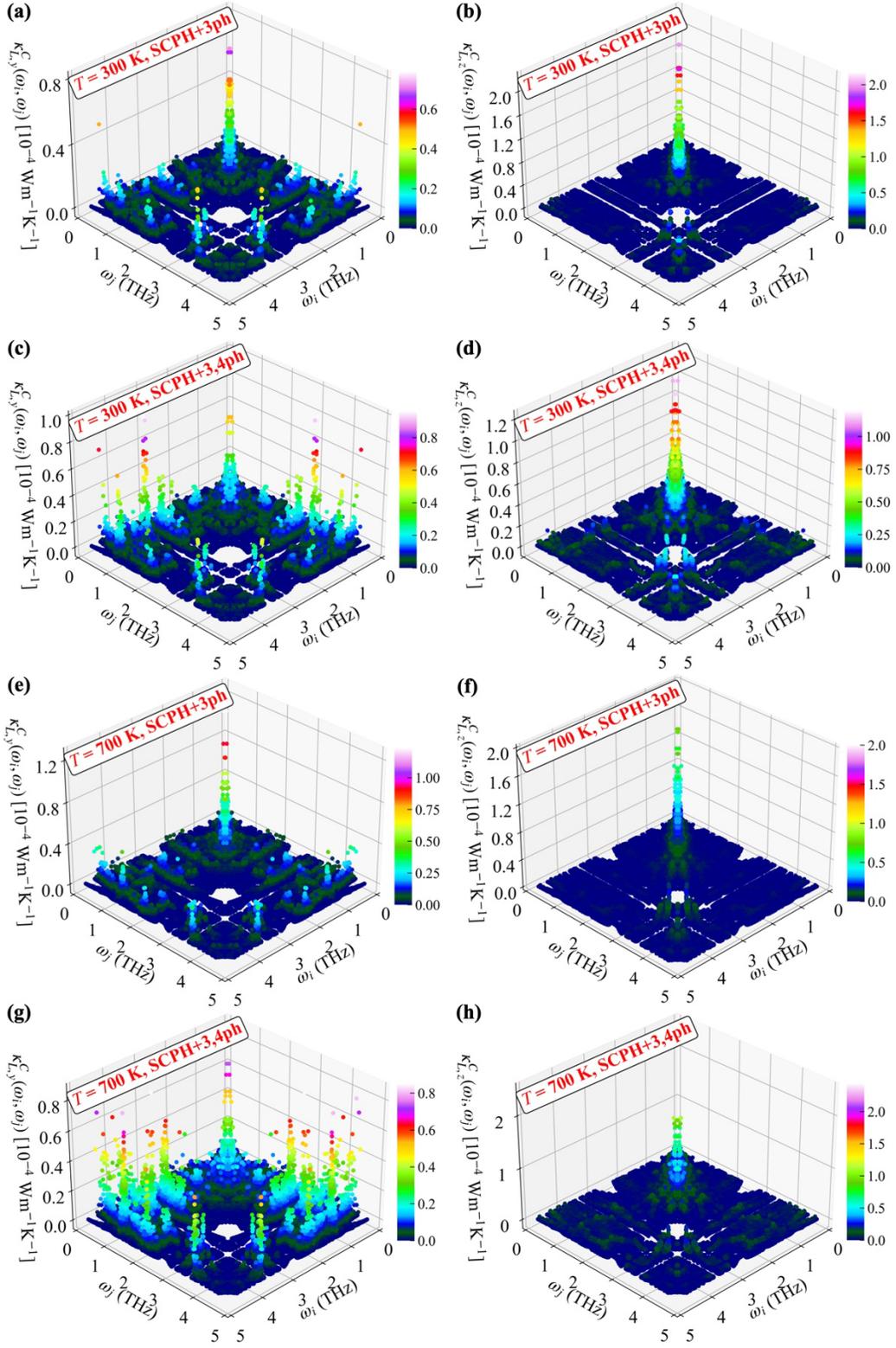

FIG. 5. Calculated Two-dimensional (2D) modal coherences' thermal conductivity $\kappa_L^C(\omega_i, \omega_j)$ obtained with (SCPH+3,4ph model) or without (SCPH + 3ph model) considering 4ph scattering rates at 300 and 700 K, respectively.

(a) Two-dimensional modal $\kappa_{L,y}^C(\omega_i, \omega_j)$ considering only 3ph scattering rates (SCPH+ 3ph model) along the y axis at 300 K. (b) The same as (a), but along the z axis. (c) Two-dimensional modal $\kappa_{L,y}^C(\omega_i, \omega_j)$ considering both 3ph and 4ph scattering rates (SCPH+3,4ph model) along the y axis at 300 K. (d) The same as (c), but along the z axis. (e) Two-dimensional modal $\kappa_{L,y}^C(\omega_i, \omega_j)$ considering only 3ph scattering rates (SCPH+ 3ph model) along the y axis at 700 K. (f) The same as (e), but along the z axis. (g) Two-dimensional modal $\kappa_{L,y}^C(\omega_i, \omega_j)$ considering both 3ph and 4ph scattering rates (SCPH+3,4ph model) along the y axis at 700 K. (h) The same as (g), but along the z axis.

To gain a deeper insight into the wave-like tunnelling of phonons in thermal transport of CsCu$_2$I$_3$, we calculate the two-dimensional modal $\kappa_L^C$ at 300 and 700 K, respectively, as illustrated in Figs. 5(a-h) and S6(a-d). The thermal conductivity $\kappa_L^C$, associated with coherences, is determined by the characteristics of two distinct phonons, evaluated through the phonon linewidth and inter-branch spacings [46,47]. In Figs. 5(a-h), when considering only 3ph scatterings, the quasi-degenerate phonon states, namely, two phonons with similar frequencies, exert a predominant influence on both the cross-chain $\kappa_{L,x}^C$ and chain-axis $\kappa_{L,z}^P$ at both 300 and 700 K. With further including 4ph scattering effects, the non-degenerate phonon states start to dominate the cross-chain $\kappa_{L,x}^C$, as illustrated in Figs. 5(c) and (g). This phenomenon can be attributed to the presence of strong 4ph scattering rates (large phonon linewidths), which promote the coupling of two phonons with significantly different frequencies [40,63]. However, the enhancement in chain-axis $\kappa_{L,z}^C$ induced by 4ph scatterings is minor and negligible, as shown in Figs. 5 (d) and (h). The varying effects of 4ph scattering rates on coherences' thermal conductivity along cross-chain and chain-axis directions can be elucidated by the wave-like nature of phonons. The remarkably low group velocities, a consequence of the chain-like structure of CsCu$_2$I$_3$, particularly in the high-frequency region, lead to a pronounced wave-like nature of phonons along cross-chain directions. This, coupled with the presence of large phonon linewidths, contributes to the significant cross-chain coherences' conductivity. To be more specific, the couplings between phonons with very different frequencies, such as 1/3.2 THz at 300 K and 1.2/3.3 THz at 700 K, make a noticeable contribution to cross-chain $\kappa_{L,x}^C$ [see Figs. 5(c, g)]. This obvious contribution is a result of the strong wave-like nature of phonons induced by the exceptionally low group velocities, as illustrated in Fig. 2(c). Conversely, the relatively high group velocities ensure the particle-like nature of phonons along the chain axis, leading to a limited population of phonons with wave-like nature.

**f) Temperature-dependent thermal conductivity**

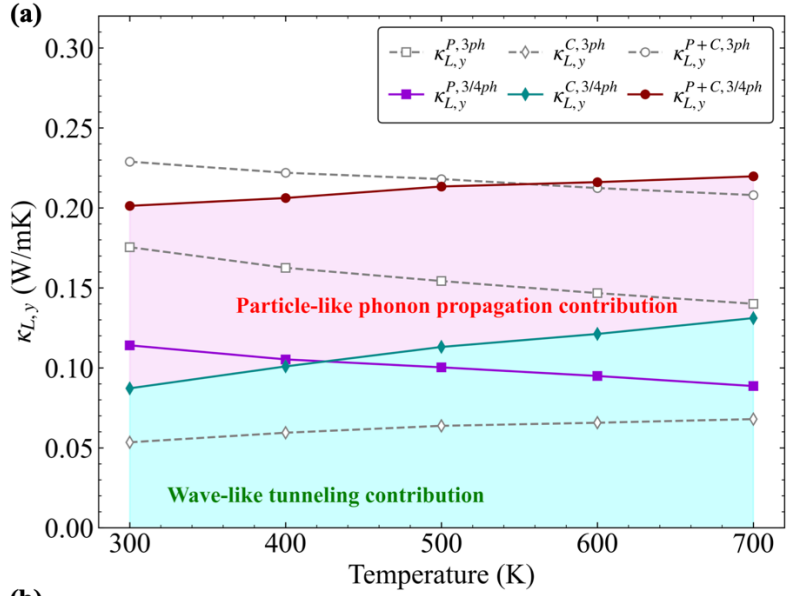

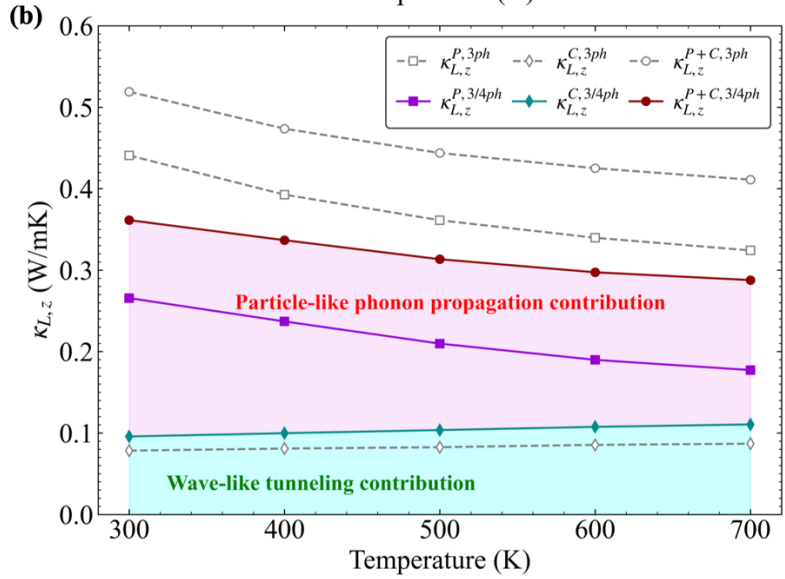

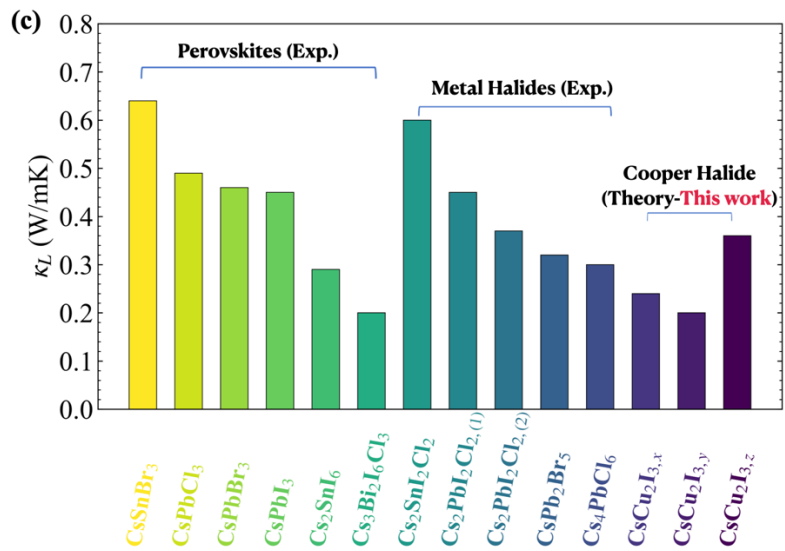

FIG. 6. (a) Calculated temperature-dependent lattice thermal conductivity $k_L$ along the y axis including populations' and coherences' contributions obtained with or without further considering four-phonon scatterings over the temperature range of 300 – 700 K. The pink and light blue shaded areas indicate the contributions from particle-like phonon propagation and wave-like tunnelling of phonons channels, respectively. (b) The same as (a), but along the z axis. (c) Comparison of our predicted lattice thermal conductivity of crystalline $CsCu_2I_3$ with other experimentally measured ultra-low thermal conductivities of inorganic perovskites and metal halides [2,4,5,22,84-87].

We next move on to examine the temperature dependence of lattice thermal conductivity in crystalline $CsCu_2I_3$ over the temperature range of 300-700 K, as illustrated in Figs. 6(a-b) and S7 in the SM [70]. When considering only 3ph scatterings, our predicted cross-chain $\kappa_{L,y}^P$ and chain-axis $\kappa_{L,z}^P$ exhibit a weak temperature dependence, roughly proportional to $\sim T^{-0.262}$ and $\sim T^{-0.367}$, respectively. The weak temperature dependences of $\kappa_{L,y}^P$ and $\kappa_{L,z}^P$ contradict the conventional temperature dependence of $\kappa_L^P$, which typically follows a $\sim T^{-1}$ relationship [88]. This phenomenon is attributed to anharmonic phonon renormalization, which is also observed materials like $BaZrO_3$ [63] and double perovskite $Cs_2AgBiBr_6$ [38]. With the inclusion of the effects of 4ph scatterings, the temperature dependence of $\kappa_{L,y}^P$ and $\kappa_{L,z}^P$ intensifies, now following $\sim T^{-0.283}$ and $\sim T^{-0.483}$, respectively [see Figs. 6(a-b)]. The enhancement in the temperature dependence of $\kappa_L^P$ can be ascribed to the increase in 4ph scattering rates with increasing temperature, which aligns with prior observations by Feng *et al.* [66] and Zheng *et al.* [63]. In general, the further incorporation of coherences' contribution $\kappa_L^C$ results in a weaker temperature dependence of the total $\kappa_L$. When considering only 3ph scatterings, the temperature dependence of cross-chain $\kappa_{L,y}^P$ changes from $\sim T^{-0.262}$ to $\sim T^{-0.111}$ and chain-axis $\kappa_{L,z}^P$ changes from $\sim T^{-0.367}$ to $\sim T^{-0.279}$. Meanwhile, both cross-chain $\kappa_{L,y}^P$ and chain-axis $\kappa_{L,z}^P$ continue to play their dominant role in the thermal transport of $CsCu_2I_3$ [see Figs. 6(a-b)]. Interestingly, the inclusion of 4ph scatterings gives rise to distinct behaviors in the chain-axis and cross-chain thermal conductivity of $CsCu_2I_3$. The total chain-axis $\kappa_{L,z}$ exhibits a relatively mild temperature dependence of $\sim T^{-0.276}$ and show a decreased trend with increasing temperature [see Fig. 6(b)]. In addition, the populations' thermal conductivity from particle-like phonons still dominates the total chain-axis thermal conductivity $\kappa_{L,z}$. In contrast, the total cross-chain thermal conductivity $\kappa_{L,y}$ displays a highly unusual temperature dependence of $\sim T^{0.106}$, implying that thermal conductivity increases with increasing temperature. This unusual increased trend of cross-chain thermal conductivity is unexpected for single crystal and can be attributed to the dominant role of cross-chain coherences' conductivity in $CsCu_2I_3$ [see Fig. 6(a)]. Considering the gentle increase in thermal conductivity with rising temperature, we classify it as an atypical glass-like behavior. From Fig. 6(a) we observe that the coherences' conductivity surpasses the populations' conductivity above 422 K. Overall, the chain-like structure of $CsCu_2I_3$, characterized by cross-chain weak bonding, leads to exceedingly low populations' conductivity and the dominant coherences' conductivity in thermal transport. Consequently, this results in the atypical trend of total cross-chain $\kappa_{L,y}$ increasing with rising temperature.

With the total $\kappa_L$ in hand, we compare the predicted $\kappa_L$ of crystalline $CsCu_2I_3$ with the experimentally measured $\kappa_L$ of perovskites and metal halides at 300 K [2,4,5,22,84-87], as illustrated in Fig. 6(c). Our predicted room-temperature total $\kappa_L$ for $CsCu_2I_3$ is 0.362 $Wm^{-1}K^{-1}$ for the chain-axis (Z-axis), 0.201 $Wm^{-1}K^{-1}$ for the cross-chain (Y-axis), and 0.239 $Wm^{-1}K^{-1}$ for the cross-chain (X-axis), as shown in Figs. 6(a-c) and S7 in the SM [70]. We note that recent theoretical findings have reported record-low lattice thermal conductivities of 0.02 $Wm^{-1}K^{-1}$ for $Cs_3Cu_2I_5$ [35] and 0.05 $Wm^{-1}K^{-1}$ for $CsCu_2I_3$ with a space group of ***Amm2*** [36]. We attribute their predicted record-low thermal conductivity to the omission of temperature effects and the neglect of coherences' contributions to thermal conductivity [40,41,46,47,63]. Our theoretical $\kappa_L$ for crystalline $CsCu_2I_3$ falls within the range of experimentally measured $\kappa_L$ for perovskites [2,4,22,84,85] and metal halides [5,84,86,87], which spans from 0.19 $Wm^{-1}K^{-1}$ to 0.64 $Wm^{-1}K^{-1}$ [see Fig. 6(c)]. The complex structure of the halide perovskite $Cs_3Bi_2I_6Cl_3$ contributes to its ultra-low thermal conductivity of 0.19 Wm$^{-1}$K$^{-1}$ [85]. Additionally, the chain-axis bond strength of CsCu$_2$I$_3$ is similar to the in-plane bond strength of Cs$_2$PbI$_2$Cl$_2$ [5]. Consequently, we expect a similar value of 0.37 Wm$^{-1}$K$^{-1}$ for the chain-axis $\kappa_{L,z}$ and lower values for the cross-chain $\kappa_{L,x,y}$, attributable to the weak cross-chain bonding in CsCu$_2$I$_3$. Overall, the predicted total $\kappa_L$ of CsCu$_2$I$_3$ in this study is reasonably reliable and can be verified by future experimental investigations.

## IV. CONCLUSIONS

In summary, we have employed a first-principles-based framework that incoporates anharmonic phonon renormalization and a unified theory of thermal transport, accounting for both 3ph and 4ph interaction processes, to investigate the thermal conductivity in CsCu$_2$I$_3$. Our results reveal that the crystalline CsCu$_2$I$_3$, with its unique 1D chain-like structure and characterized by a ***cmcm*** space group, exhibits unstable soft modes dominated by Cu and I atoms at 0 K. Utilizing the anharmonic phonon renormalization technique, we successfully stabilize the soft modes of CsCu$_2$I$_3$ at ~ 75 K. The relatively low phase transition temperature provides an explanation for the experimentally observed CsCu$_2$I$_3$ structure, characterized by a ***cmcm*** space group at room temperature. By employing a unified theory of thermal transport that considers both the coherences and populations' contributions, we predict a room-temperature $\kappa_L$ of CsCu$_2$I$_3$ to be 0.362 Wm$^{-1}$K$^{-1}$ along the chain axis (Z-axis), 0.201 Wm$^{-1}$K$^{-1}$ along the cross-chain direction (Y-axis), and 0.239 Wm$^{-1}$K$^{-1}$ along the cross-chain direction (X-axis). Meanwhile, the contributions from populations dominate the total chain-axis thermal conductivity, while the coherences' contributions start to take the lead in the total cross-chain thermal conductivity above 442 K. The dominance of $\kappa_{L,y}^C$ in the cross-chain thermal conductivity of CsCu$_2$I$_3$ can be traced back to its unique chain-like 1D structure. This unique structure, characterized by weak bonding, results in an exceedingly low cross-chain $\kappa_{L,y}^P$, exceptionally small cross-chain group velocities, and strong 4ph scatterings, all of which collectively facilitate the wave-like tunnelling of phonons. Furthermore, our predictions unveil an unexpected anomalous trend of increasing cross-chain thermal conductivity with increasing temperature across the entire temperature range, which is atypical for a single crystal. We categorize this phenomenon as an abnormal glass-like behavior of thermal conductivity and attribute it to the dominant role of cross-chain coherences' conductivity in thermal transport of $CsCu_2I_3$. Our study emphasizes that the conventional Peierls-Boltzmann transport equation falls short in elucidating the thermal transport in crystalline $CsCu_2I_3$. It further suggests that a comprehensive understanding of thermal transport in $CsCu_2I_3$ can be achieved by considering both 3ph and 4ph scattering effects on phonon energies, the particle-like phonon propagation, and the wave-like tunnelling of phonons. Finally, our work not only offers insights into the nature of lattice thermal transport in 1D metal halides but also suggests a potential new avenue for engineering phonon-related properties.

## ACKNOWLEDGEMENT

We are thankful for the financial support from the Science and Technology Planning Project of Guangdong Province, China (Grant No. 2017A050506053), the Science and Technology Program of Guangzhou (No. 201704030107), and the Hong Kong General Research Fund (Grants No. 16214217 and No. 16206020). This paper was supported in part by the Project of Hetao Shenzhen-Hong Kong Science and Technology Innovation Cooperation Zone (HZQB-KCZYB2020083). R.G. acknowledges support from the Excellent Young Scientists Fund (Overseas) of Shandong Province (2022HWYQ091) and the Initiative Research Fund of Shandong Institute of Advanced Technology (2020107R03). C.L. acknowledges the support from the Sinergia project of the Swiss National Science Foundation (grant number CRSII5_189924).